\documentclass[
    pre,
    aps,
    amsmath,
    amssymb,
    reprint,
    superscriptaddress,
    nofootinbib
]{revtex4-2}

\usepackage{graphicx}
\usepackage{dcolumn}
\usepackage{bm}
\usepackage[
    bookmarks=false,
    colorlinks=true,
    linkcolor=blue,
    citecolor=blue,
    urlcolor=blue,
]{hyperref}
\usepackage{orcidlink}
\usepackage{titlesec}

\DeclareMathOperator{\erf}{erf}

\DeclareMathOperator{\sinc}{sinc}
\DeclareMathOperator{\sinhc}{sinhc}

\newcommand{\sparallel}{\mathrel{\raisebox{0.1ex}{\scalebox{0.5}{$\parallel$}}}}

\definecolor{aesthetic-background}{RGB}{30, 20, 40}
\definecolor{aesthetic-blue}{RGB}{0, 174, 255}
\definecolor{aesthetic-cyan}{RGB}{90, 200, 230}
\definecolor{aesthetic-green}{RGB}{55, 196, 55}
\definecolor{aesthetic-magenta}{RGB}{249, 42, 173}
\definecolor{aesthetic-yellow}{RGB}{253, 163, 42}
\definecolor{ppt-blue}{RGB}{2, 83, 118}

\makeatletter
\newcommand*\@secondofsix[6]{#2}
\newcommand{\addtotitleformat}{%
  \@ifstar{\addtotitleformat@star}{\addtotitleformat@nostar}}
\newcommand\addtotitleformat@nostar[2]{%
  \PackageError{titlesec}{non starred form of \string\addtotitleformat\space not supported}{}}
\newcommand\addtotitleformat@star[2]{%
  \expandafter\expandafter\expandafter\expandafter
  \expandafter\expandafter\expandafter\def
  \expandafter\expandafter\expandafter\expandafter
  \expandafter\expandafter\expandafter\@currentsection@font
  \expandafter\expandafter\expandafter\expandafter
  \expandafter\expandafter\expandafter{%
    \expandafter\expandafter\expandafter\@secondofsix
       \csname ttlf@\expandafter\@gobble\string#1\endcsname}%
  \titleformat*{#1}{\@currentsection@font#2}%
}
\makeatother

\addtotitleformat*{\section}{\MakeUppercase}

\makeatletter
\renewcommand*\date[1][\Dated@name]{
    \def\@date{
        #1\today
    }
}
\makeatother

\begin{document}


\title{
%
    Link length and energy fluctuations in extensible freely jointed chains
}
\author{
    Michael R. Buche%
    \:\orcidlink{0000-0003-1892-0502}\,
}
\email{mrbuche@sandia.gov}
\affiliation{
    Materials and Failure Modeling, Sandia National Laboratories, Albuquerque, New Mexico 87185, USA
}
\date{}

\begin{abstract}
The freely jointed chain is often applied to model the thermodynamics of single polymer chains, but the traditional formulation of the model lacks internal energy changes due to bond stretching.
For this reason, the extensible freely jointed chain model includes a potential energy function, typically harmonic, that governs the length of each link in the chain.
Among the other quantities of interest that are subject to thermal fluctuations, these link lengths and energies too fluctuate about their ensemble average values.
Since a plethora of models for polymer chains and networks incorporate chain dissociation as a function of either link length or energy, these fluctuations are crucial to understand and quantify.
Motivated by this fact, fluctuations in link length and energy are analyzed within a freely jointed chain under an applied force.
These fluctuations are quantified through their average values, standard deviations, and probability distributions.
Across all values, asymptotically correct analytic relations and their less ergonomic exact counterparts are introduced.
The asymptotic relations are verified to be accurate through direct comparison and to be correct within transcendentally small terms through error analysis.
In certain cases, the fluctuations are shown to be approximately normally distributed.
Hereafter, model components predicated on link length or energy ought to account for these fluctuations.
\end{abstract}

\maketitle


\section{Introduction}\label{sec:introduction}

The extensible freely jointed chain comprises a series of stretchable links joined by freely rotating hinges \cite{buche2022freely}.
Some potential energy governs the link stretches, which is typically the harmonic potential \cite{balabaev2009extension,fiasconaro2019analytical,buche2026thermodynamic}.
Under an applied force \cite{treloar1949physics}, resistance to extension of the chain is dominated by the reluctance to reduce entropy.
As the applied force becomes sufficiently large and the chain approaches full extension, the resistance to further extension is dominated by the stiffness of the links \cite{buche2021fundamental}.
Landmark single-molecule stretching experiments, including DNA overstretching \cite{smith1996overstretching} and polysaccharide force spectroscopy \cite{rief1997single}, further motivated extensible chain models that capture energetic bond stretching at large force.
Experiments like these have also demonstrated the importance of thermal fluctuations in single-molecule responses \cite{ritort2006single}.
In the isotensional ensemble \cite{manca2012elasticity}, all quantities of interest apart from the ensemble fixtures (the number of links, applied force, and temperature) are subject to thermal fluctuations \cite{mcquarrie}.
This includes each link conformation as well as the overall chain extension \cite{buche2026thermodynamic}, but also each link length and corresponding link energy.
These fluctuations are important to study in order to build a more basic understanding of the statistical thermodynamics of single polymer chains \cite{mao2017rupture,buche2022freely,mulderrig2023statistical}.
They could also be crucial for polymer network models that include chain dissociation as a function of link length or energy \cite{lavoie2019modeling,buche2021chain,mulderrig2021affine,lamont2021rate}, by enabling such models to better represent the underlying stochastic physics.
To that end, fluctuations in both link length and energy are analyzed in detail through their averages, deviations, and probability distributions.
New asymptotic relations are derived and shown to be highly accurate in comparison with exact analytic solutions, both of which are implemented in \texttt{conspire} \cite{conspire}.
Fluctuations in link length are shown to be mostly normally distributed, and the significance of neglecting these fluctuations when predicting chain dissociation is briefly demonstrated.

\section{Link length}\label{sec:length}

The statistics of each link in a freely jointed chain in the isotensional ensemble are independent and identical, as described by the single-link partition function \cite{buche2022freely}
\begin{equation}\label{eq:z:integral}
z(\eta) = \ell_b^3 \iiint e^{\eta\lambda\cos\theta - \upsilon(\lambda)}\,\lambda^2\,d\lambda\,\sin\theta\,d\theta\,d\phi.
\end{equation}
Note that $\lambda=\ell/\ell_b$ is the nondimensional link length, $\eta=\beta f\ell_b$ is the nondimensional force, $\kappa=\beta k_b\ell_b^2$ is the nondimensional link stiffness, and $\upsilon=\kappa(\lambda-1)^2/2$ is the assumed harmonic nondimensional link energy function.
For $\kappa\gg 1$, Eq.~\eqref{eq:z:integral} is well-approximated by \cite{fiasconaro2019analytical,buche2022freely}
\begin{equation}\label{eq:z:asymptotic}
z(\eta) \sim 4\pi\ell_b^3\sqrt{\frac{2\pi}{\kappa}}\sinhc(\eta)\,e^{\eta^2/2\kappa}\left[1 + \frac{\eta}{\kappa}\,\coth(\eta)\right],
\end{equation}
where $\sinhc$ is the hyperbolic $\sinc$ function.
Notably, Eq.~\eqref{eq:z:asymptotic} is correct within transcendentally small terms in certain regimes when compared to the exact result \cite{balabaev2009extension,buche2022freely}
\begin{equation}\label{eq:z:exact}
z(\eta) = \pi\ell_b^3\sqrt{\frac{2\pi}{\kappa}}\,\frac{e^{\eta^2/2\kappa}}{\eta}\left[\mu_0^+(\kappa,\eta) - \mu_0^-(\kappa,\eta)\right],
\end{equation}
where the functions $\mu_0^+(\kappa,\eta)$ are defined in Eq.~\eqref{eq:mu:pm}.
Similarly accurate asymptotic relations will be derived here for the moments and distribution of link lengths.
One particularly useful relation is the ensemble average,
\begin{equation}\label{eq:ensemble:average}
\langle f\rangle = \frac{4\pi\ell_b^3}{z(\eta)}\int_0^\infty f(\lambda)\,\sinhc(\eta\lambda)\,e^{-\upsilon(\lambda)}\,\lambda^2\,d\lambda,
\end{equation}
where $f(\lambda)$ is any function of $\lambda$, such as $f(\lambda) = \lambda^n$.

\subsection{Link length moments}\label{sec:length:moments}

\begin{figure}[t]
\includegraphics{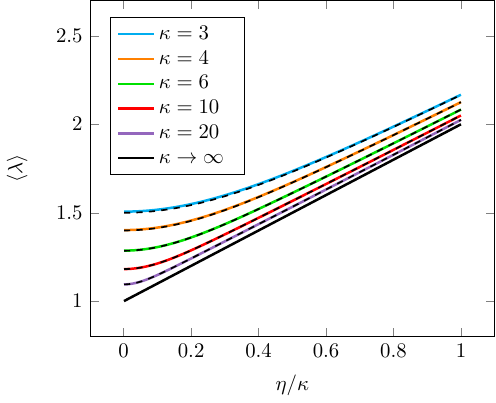}
\caption{\label{fig:length:average}%
Ensemble average $\langle\lambda\rangle$ of the nondimensional link length $\lambda$ as a function of the nondimensional force $\eta$ using exact (solid) and asymptotic (dashed) relations for increasing nondimensional link stiffness $\kappa$.
}
\end{figure}

For $f(\lambda) = \lambda^n$, Eq.~\eqref{eq:ensemble:average} gives $\langle\lambda^n\rangle=z_n/z_0$, where
\begin{equation}\label{eq:z:n}
z_n(\eta) \equiv 4\pi\ell_b^3\int_0^\infty \lambda^n \sinhc(\eta\lambda)\,e^{-\upsilon(\lambda)}\,\lambda^2\,d\lambda.
\end{equation}
When $\kappa\gg 1$, the nondimensional potential $\upsilon(\lambda)$ is steep and Eq.~\eqref{eq:z:n} could be asymptotically approximated \cite{buche2021fundamental}.
Previous expansions were centered about the minimum of the steep potential energy \cite{buche2022freely,buche2023modeling,buche2024statistical}, here located at $\hat{\lambda}=1$.
Now that the desired asymptotic relation is the argument of the steep potential itself, it becomes more appropriate to center the expansion about the minimizer of the total potential energy $\upsilon-\eta\lambda$, which would be $\hat{\lambda}=1+\eta/\kappa$.
Laplace's method \cite{bender2013advanced} applied to Eq.~\eqref{eq:z:n} then yields
\begin{equation}
z_n(\eta) \sim 2\pi\ell_b^3\sqrt{\frac{2\pi}{\kappa}}\,\frac{e^\eta}{\eta}\,e^{\eta^2/2\kappa} \left[g_n(\hat{\lambda}) + \frac{g_n''(\hat{\lambda})}{2\kappa}\right],
\end{equation}
where $g_n(\lambda)\equiv\lambda^{n+1}(1-e^{-2\eta\lambda})$.
The ratio of $z_n$ to $z_0$ is written and the fraction subsequently expanded as
\begin{equation}\label{eq:z:n:0}
\frac{z_n}{z_0} \sim \frac{g_n + g_n''/2\kappa}{g_0 + g_0''/2\kappa} \sim \frac{g_n}{g_0} + \frac{1}{2\kappa}\left[\frac{g_n''}{g_0} - \frac{g_n g_0''}{g_0^2}\right],
\end{equation}
which equals $\langle\lambda^n\rangle$ via Eqs.~\eqref{eq:ensemble:average} and \eqref{eq:z:n}.
With $g_{n+1}=\lambda g_n$, Eq.~\eqref{eq:z:n:0} can then be rewritten as simply
\begin{equation}\label{eq:lambda:n}
\langle\lambda^n\rangle \sim \hat{\lambda}^n + \frac{\hat{\lambda}^{n-2}}{2\kappa}\left[n(n-1) + 2n\hat{\lambda}\,\frac{g_0'(\hat{\lambda})}{g_0(\hat{\lambda})}\right],
\end{equation}
so with $n=1$, the average link stretch is approximately
\begin{equation}
\langle\lambda\rangle \sim \hat{\lambda} + \frac{1}{\kappa}\left[\frac{1}{\hat{\lambda}} + \eta\left(\coth(\hat{\lambda}\eta) - 1\right)\right].
\end{equation}
Now the saddle point location $\hat{\lambda} = 1+\eta/\kappa$ is applied, taking $\coth(\hat{\lambda}\eta)\sim\coth(\eta)$ and collecting terms above the common denominator $1+(\eta/\kappa)\coth(\eta)$ from Eq.~\eqref{eq:z:asymptotic}.
The resulting asymptotic approximation of the ensemble average nondimensional link length for $\kappa\gg 1$ is
\begin{equation}\label{eq:length:average:asymptotic}
\langle\lambda\rangle \sim 1 + \frac{h^\lambda_1(\kappa,\eta)}{1 + (\eta/\kappa)\coth(\eta)} + \frac{\eta}{\kappa},
\end{equation}
where $h^\lambda_1(\kappa,\eta) \equiv 1/\kappa + (\eta/\kappa)(1 - \eta/\kappa)(\coth\eta - 1)$.
Eq.~\eqref{eq:length:average:asymptotic} is shown in Fig.~\ref{fig:length:average} as a function of the scaled nondimensional force $\eta/\kappa$ along with the exact relation from Eq.~\eqref{eq:lambda:1:exact}.
The asymptotic relation maintains a high degree of accuracy across the range of $\eta$ and different $\kappa$.
The limit $\langle\lambda\rangle\to 1+\eta/\kappa$ as $\kappa\to\infty$ is also shown.

\begin{figure}[t]
\includegraphics{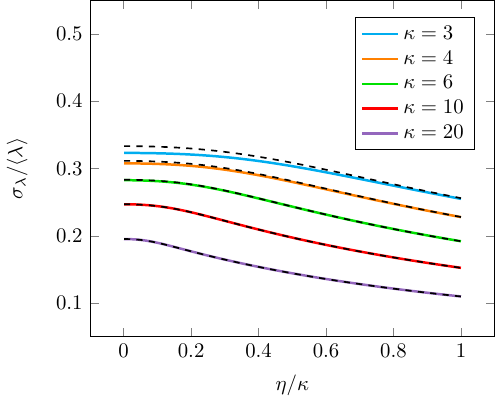}
\caption{\label{fig:length:deviation}%
Standard deviation $\sigma_\lambda$ of the nondimensional link length $\lambda$ as a function of the nondimensional force $\eta$ using exact (solid) and asymptotic (dashed) relations for increasing nondimensional link stiffness $\kappa$.
}
\end{figure}

Taking $n=2$ in Eq.~\eqref{eq:lambda:n}, the average squared nondimensional link length is approximately
\begin{equation}
\langle\lambda^2\rangle \sim \hat{\lambda}^2 + \frac{1}{\kappa}\left[3 + 2\eta\hat{\lambda}\left(\coth(\hat{\lambda}\eta) - 1\right)\right].
\end{equation}
The saddle point location $\hat{\lambda} = 1+\eta/\kappa$ is applied again, making the same manipulations as previously to collect terms above the common denominator $1+(\eta/\kappa)\coth(\eta)$.
The resulting asymptotic approximation of the variance $\sigma_\lambda^2 = \langle\lambda^2\rangle - \langle\lambda\rangle^2$ valid for $\kappa\gg 1$ is
\begin{equation}\label{eq:length:variance:asymptotic}
\sigma_\lambda^2 \sim 1 + \frac{h^\lambda_2(\kappa,\eta)}{1 + (\eta/\kappa)\coth(\eta)} + \frac{\eta^2}{\kappa^2} - \langle\lambda\rangle^2,
\end{equation}
where $h^\lambda_2(\kappa,\eta) \equiv 3/\kappa + 2\eta^2/\kappa^2 + (3/\kappa + 2)(\eta/\kappa)\coth\eta$.
Eq.~\eqref{eq:length:variance:asymptotic} is shown in Fig.~\ref{fig:length:deviation} as the relative standard deviation $\sigma_\lambda/\langle\lambda\rangle$ as a function of the scaled nondimensional force $\eta/\kappa$ along with the exact relation using Eq.~\eqref{eq:lambda:2:exact}.
This result is similarly accurate across $\eta$ and $\kappa$.

\subsection{Link length distribution}\label{sec:length:distribution}

\begin{figure}[t]
\includegraphics{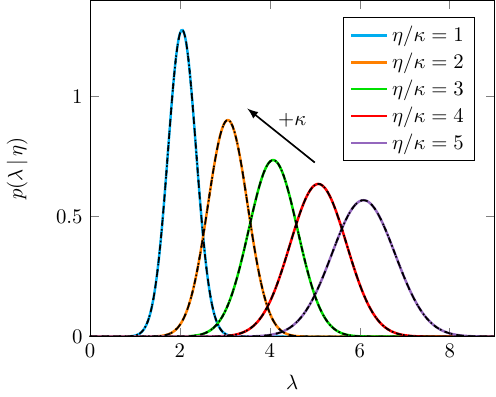}
\caption{\label{fig:length:probability-fixed-eta}%
Probability density $p$ of the nondimensional link length $\lambda$ given nondimensional force $\eta$ using exact (solid), asymptotic (dashed), and Gaussian (dotted) relations while increasing nondimensional link stiffness $\kappa$ with fixed $\eta=10$.
}
\end{figure}

The probability density distribution of nondimensional link lengths $\lambda$ given the nondimensional force $\eta$ can be written as the ensemble average via Eq.~\eqref{eq:ensemble:average}
\begin{equation}\label{eq:length:probability:definition}
p(\lambda\,|\,\eta) = \big\langle\delta(\lambda' - \lambda)\big\rangle.
\end{equation}
Since $\lambda$ is the variable of integration, this probability is simply the integrand of Eq.~\eqref{eq:ensemble:average} scaled by the partition function $z(\eta)$.
Utilizing the relation for $z(\eta)$ in Eq.~\eqref{eq:z:asymptotic}, the asymptotic approximation of the probability is then
\begin{equation}\label{eq:length:probability:asymptotic}
p(\lambda\,|\,\eta) \sim \sqrt{\frac{\kappa}{2\pi}}\,\frac{\mathrm{sinhc}(\eta\lambda)}{\mathrm{sinhc}(\eta)}\,\frac{\lambda^2\,e^{-\upsilon(\lambda)}\,e^{-\eta^2/2\kappa}}{1 + (\eta/\kappa)\coth(\eta)}.
\end{equation}
Analogously, the exact relation for $z(\eta)$ in Eq.~\eqref{eq:z:exact} allows the exact relation for the probability to be written as
\begin{equation}\label{eq:length:probability:exact}
p(\lambda\,|\,\eta) = \sqrt{\frac{\kappa}{2\pi}}\,\frac{4\lambda\sinh(\eta\lambda)\,e^{-\upsilon(\lambda)}\,e^{-\eta^2/2\kappa}}{\mu_0^+(\kappa,\eta) - \mu_0^-(\kappa,\eta)},
\end{equation}
where $\mu_0^\pm(\kappa,\eta)$ is defined by Eq.~\eqref{eq:mu:pm}.
The asymptotic relation in Eq.~\eqref{eq:length:probability:asymptotic} is shown in Fig.~\ref{fig:length:probability-fixed-eta} as a function of the nondimensional link length $\lambda$ along with the exact relation using Eq.~\eqref{eq:length:probability:exact}, where $\eta=10$ is fixed while $\kappa$ varies between the different curves.
Also shown in Fig.~\ref{fig:length:probability-fixed-eta} are the Gaussian approximations via the asymptotic relations for the average and variance in Eqs.~\eqref{eq:length:average:asymptotic} and \eqref{eq:length:variance:asymptotic}.
Each of the three cases -- the exact relation, asymptotic approximation, and Gaussian approximation -- are wholly indistinguishable from one another in Fig.~\ref{fig:length:probability-fixed-eta}, demonstrating the expected accuracy of not only the asymptotic theory, but also the accuracy of a normally-distributed approximation in this particular case.

The asymptotic relation in Eq.~\eqref{eq:length:probability:asymptotic} is shown again in Fig.~\ref{fig:length:probability-fixed-kappa} as a function of the nondimensional link length $\lambda$ along with the exact relation using Eq.~\eqref{eq:length:probability:exact}, where $\kappa=10$ is fixed while $\eta$ varies between different curves.
Gaussian approximations are also shown again via the asymptotic relations for the average and variance in Eqs.~\eqref{eq:length:average:asymptotic} and \eqref{eq:length:variance:asymptotic}.
Once again, all three relations are indistinguishable from one another once plotted.

\subsection{Discussion}\label{sec:length:discussion}

\begin{figure}[t]
\includegraphics{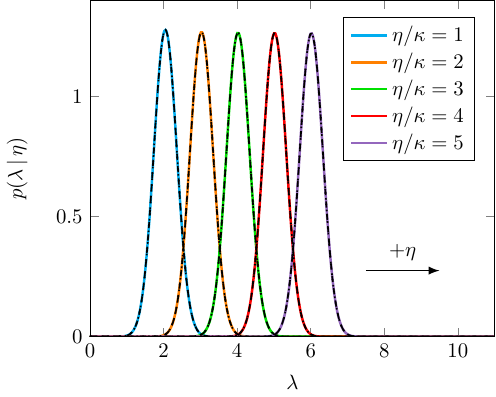}
\caption{\label{fig:length:probability-fixed-kappa}%
Probability density $p$ of the nondimensional link length $\lambda$ given nondimensional force $\eta$ using exact (solid), asymptotic (dashed), and Gaussian (dotted) relations while increasing $\eta$ with fixed nondimensional link stiffness $\kappa=10$.
}
\end{figure}

The results in Fig.~\ref{fig:length:average} show that without an applied force, the average link stretch is generally above unity.
Nonzero temperatures give rise to thermal fluctuations (see also the nonzero deviation at zero force in Fig.~\ref{fig:length:deviation}), and the tendency toward higher entropy at equilibrium causes the average to move above unity (where there is more phase space volume) to approximately $1+2/(\kappa+1)$.
The relative fluctuations in Fig.~\ref{fig:length:deviation} decrease with the applied force, but most of this is due to the average decreasing with force.
For $\kappa>1$, the absolute variance $\sigma_\lambda^2$ is nearly constant at $1/\kappa$ for any $\eta$.
Further, other moments of the link length distribution appear to be negligible for $\kappa>1$, as evidenced by the direct overlap of the Gaussian distribution with the exact (and asymptotic) results in Figs.~\ref{fig:length:probability-fixed-eta} and \ref{fig:length:probability-fixed-kappa}.
As $\kappa$ increases for fixed $\eta$, Fig.~\ref{fig:length:probability-fixed-eta} shows the distribution narrowing and deforming toward $\langle\lambda\rangle\to 1$.
As $\eta$ increases for fixed $\kappa$, Fig.~\ref{fig:length:probability-fixed-kappa} shows the distribution simply translating via $\langle\lambda\rangle\sim 1+\eta/\kappa$ with approximately constant $\sigma_\lambda^2\sim 1/\kappa$.
A normal distribution is accurate because, for $\kappa\gg 1$, the probability in Eq.~\eqref{eq:length:probability:asymptotic} localizes near $\langle\lambda\rangle$ with small width $O(\kappa^{-1/2})$, while remaining non-Gaussian factors, such as $\lambda^2$ and $\sinhc(\eta\lambda)$, vary slowly over that narrow region.

\section{Link energy}\label{sec:energy}

\begin{figure}[t]
\includegraphics{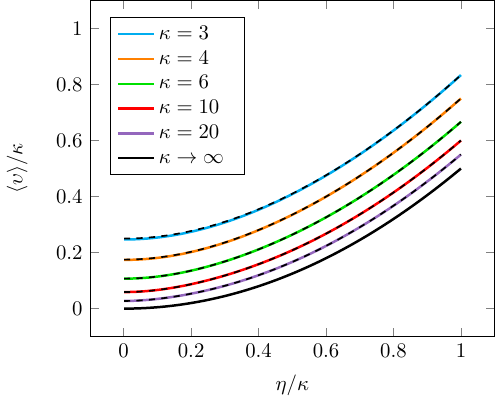}
\caption{\label{fig:energy:average}%
Ensemble average $\langle\upsilon\rangle$ of the nondimensional link energy $\upsilon$ as a function of the nondimensional force $\eta$ using exact (solid) and asymptotic (dashed) relations for increasing nondimensional link stiffness $\kappa$.
}
\end{figure}

Highly accurate asymptotic relations for the average, variance, and probability distribution of the nondimensional link energy $\upsilon$ are now derived.
Important to this derivation is the nondimensional link free energy, defined here as $\varrho(\eta) = -\ln z(\eta)$.
Average values and other moments of the internal energy are calculated using temperature derivatives \cite{mcquarrie}.
With $\beta=1/kT$, the ensemble average nondimensional total potential energy is
\begin{equation}\label{eq:energy:total:pe}
\beta\,\frac{\partial\varrho}{\partial\beta} = \langle\upsilon\rangle - \eta\langle\gamma^{\sparallel}\rangle,
\end{equation}
where $\langle\gamma^{\sparallel}\rangle = \langle\lambda\cos\theta\rangle$ is the ensemble average of the nondimensional longitudinal extension for the link \cite{buche2026thermodynamic}.
Note that any energy of an extensible freely jointed chain in the isotensional ensemble is simply the corresponding energy of a link multiplied by the number of links $N_b$ in the chain, since the links are independent and identical for this particular combination of model and ensemble.
Here, only the moments of the link potential energy $\upsilon$ are required, which can be calculated quite easily through taking derivatives of the nondimensional link stiffness $\kappa$.
The ensemble average of the link energy is
\begin{equation}\label{eq:energy:average:definition}
\langle\upsilon\rangle = \kappa\,\frac{\partial\varrho}{\partial\kappa},
\end{equation}
and the variance of the link energy $\sigma_\upsilon^2 = \langle\upsilon^2\rangle - \langle\upsilon\rangle^2$ is
\begin{equation}\label{eq:energy:variance:definition}
\sigma_\upsilon^2 = -\kappa^2\,\frac{\partial^2\varrho}{\partial\kappa^2},
\end{equation}
which can be verified using Eqs.~\eqref{eq:z:integral} and \eqref{eq:ensemble:average}.
Note that these relations are still valid for arbitrary link potentials by using $\kappa$ or any other parameter in place of it which only appears in Eq.~\eqref{eq:z:integral} in proportion to $\upsilon$.

\subsection{Link energy moments}\label{sec:energy:moments}

\begin{figure}[t]
\includegraphics{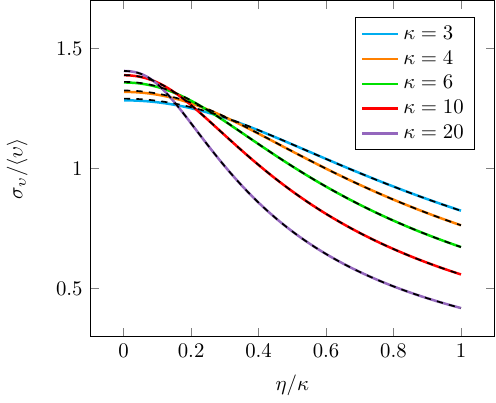}
\caption{\label{fig:energy:deviation}%
Standard deviation $\sigma_\upsilon$ of the nondimensional link energy $\upsilon$ as a function of the nondimensional force $\eta$ using exact (solid) and asymptotic (dashed) relations for increasing nondimensional link stiffness $\kappa$.
}
\end{figure}

Back in Sec.~\ref{sec:length:moments}, novel asymptotic expansions about a moving minimum were introduced since the quantity of interest $\lambda$ was the argument of the steep potential $\upsilon$.
That is not the case here, the quantity of interest is the steep potential $\upsilon$ itself, but the original asymptotic expansion \cite{buche2022freely} in Eq.~\eqref{eq:z:asymptotic} can be used.
Applying Eq.~\eqref{eq:energy:average:definition}, the asymptotic approximation of the ensemble average nondimensional link energy for $\kappa\gg 1$ is
\begin{equation}\label{eq:energy:average:asymptotic}
\langle\upsilon\rangle \sim \frac{1}{2} + \frac{h^\upsilon_1(\kappa,\eta)}{1 + (\eta/\kappa)\coth(\eta)} + \frac{\eta^2}{2\kappa},
\end{equation}
where $h^\upsilon_1(\kappa,\eta) \equiv (\eta/\kappa)\coth(\eta)$.
Eq.~\eqref{eq:energy:average:asymptotic} is shown in Fig.~\ref{fig:energy:average} to be quite accurate compared to the exact relation
\begin{equation}\label{eq:energy:average:exact}
\langle\upsilon\rangle = \frac{\kappa}{2}\Big(\langle\lambda^2\rangle - 2\langle\lambda\rangle + 1\Big),
\end{equation}
where $\langle\lambda\rangle$ and $\langle\lambda^2\rangle$ are given by Eqs.~\eqref{eq:lambda:1:exact} and \eqref{eq:lambda:2:exact}.
The limit $\langle\upsilon\rangle/\kappa\to\eta^2/2\kappa^2$ as $\kappa\to\infty$ is also shown in Fig.~\ref{fig:energy:average}.
Similarly applying Eq.~\eqref{eq:energy:variance:definition} to Eq.~\eqref{eq:z:asymptotic}, one obtains
\begin{equation}\label{eq:energy:variance:asymptotic}
\sigma_\upsilon^2 \sim \frac{1}{2} + \frac{h^\upsilon_2(\kappa,\eta)}{1 + (\eta/\kappa)\coth(\eta)} + \frac{\eta^2}{\kappa},
\end{equation}
where $h^\upsilon_2(\kappa,\eta) \equiv h^\upsilon_1(\kappa,\eta)(2-h^\upsilon_1(\kappa,\eta)/(1 + (\eta/\kappa)\coth\eta))$.
As the relative fluctuation $\sigma_\upsilon/\langle\upsilon\rangle$, Eq.~\eqref{eq:energy:variance:asymptotic} is shown in Fig.~\ref{fig:energy:deviation} to accurately approximate the exact result
\begin{equation}\label{eq:energy:variance:exact}
\sigma_\upsilon^2 = \frac{\kappa^2}{4}\Big(\langle\lambda^4\rangle - 4\langle\lambda^3\rangle + 6\langle\lambda^2\rangle - 4\langle\lambda\rangle + 1\Big) - \langle\upsilon\rangle^2,
\end{equation}
where $\langle\lambda\rangle$ through $\langle\lambda^4\rangle$ are given by Eqs.~\eqref{eq:lambda:1:exact}--\eqref{eq:lambda:4:exact}.

\subsection{Link energy distribution}\label{sec:energy:distribution}

\begin{figure}[t]
\includegraphics{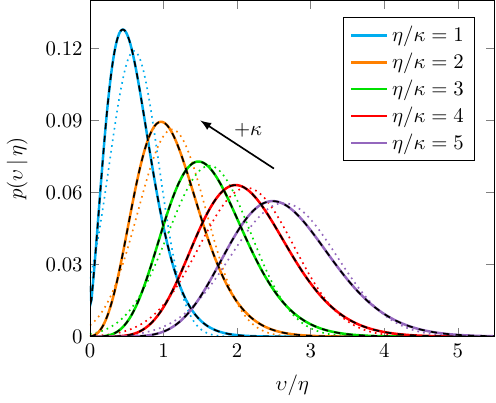}
\caption{\label{fig:energy:probability-fixed-eta}%
Probability density $p$ of the nondimensional link energy $\upsilon$ given nondimensional force $\eta$ using exact (solid), asymptotic (dashed), and Gaussian (dotted) relations while increasing nondimensional link stiffness $\kappa$ with fixed $\eta=10$.
}
\end{figure}

The probability density distribution of nondimensional link energies $\upsilon$ given the nondimensional force $\eta$ can be written as the ensemble average via Eq.~\eqref{eq:ensemble:average}
\begin{equation}\label{eq:energy:probability:definition}
p(\upsilon\,|\,\eta) = \big\langle\delta(\upsilon(\lambda') - \upsilon)\big\rangle.
\end{equation}
The Dirac delta function in effect plucks out only the nondimensional link lengths $\lambda'$ which correspond to the nondimensional link energy $\upsilon$ when integrating over each configuration in Eq.~\eqref{eq:ensemble:average}.
In other words, if an inverse function $\lambda(\upsilon)$ exists for a given link potential $\upsilon(\lambda)$, 
\begin{equation}\label{eq:energy:probability:integral}
p(\upsilon\,|\,\eta) = \int_0^\infty p(\lambda(\upsilon)\,|\,\eta) \,d\lambda.
\end{equation}
In practice this integral becomes a sum, and for the harmonic link potential $\upsilon(\lambda)=\kappa(\lambda - 1)^2/2$, this sum has two terms corresponding to the roots $\lambda^\pm\equiv 1\pm\sqrt{2\upsilon/\kappa}$.
A change of variables is also required, i.e.,
\begin{equation}\label{eq:energy:probability:jacobian}
d\lambda = \frac{\partial\lambda}{\partial\upsilon}\,d\upsilon = \frac{1}{\eta(\upsilon)}\,d\upsilon,
\end{equation}
where $\eta(\upsilon) = \sqrt{2\kappa\upsilon}$ for the harmonic potential.
Finally, the probability density distribution of nondimensional link energies can be written in terms of that for the nondimensional link lengths as
\begin{equation}\label{eq:energy:probability:exact}
p(\upsilon\,|\,\eta) = \frac{p(\lambda^+\,|\,\eta) + p(\lambda^-\,|\,\eta)}{\sqrt{2\kappa\upsilon}}.
\end{equation}
Eq.~\eqref{eq:energy:probability:exact} is evaluated using either the asymptotic relation for $p(\lambda\,|\,\eta)$ in Eq.~\eqref{eq:length:probability:asymptotic} or the exact relation in Eq.~\eqref{eq:length:probability:exact}, for fixed $\eta=10$ and varying $\kappa$.
The results in Fig.~\ref{fig:energy:probability-fixed-eta} show that the asymptotic approximation is highly accurate.
The Gaussian distributions via the asymptotic relations for the average and variance in Eqs.~\eqref{eq:energy:average:asymptotic} and \eqref{eq:energy:variance:asymptotic} are also shown in Fig.~\ref{fig:energy:probability-fixed-eta}, which are inaccurate approximations.

Eq.~\eqref{eq:energy:probability:exact} is again evaluated using Eq.~\eqref{eq:length:probability:asymptotic} or Eq.~\eqref{eq:length:probability:exact}, this time for fixed $\kappa=10$ and varying $\eta$, along with the Gaussian distributions via Eqs.~\eqref{eq:energy:average:asymptotic} and \eqref{eq:energy:variance:asymptotic}, with the results shown in Fig.~\ref{fig:energy:probability-fixed-kappa}.
The asymptotic approximation is again highly accurate, while the Gaussian distribution seems to become somewhat accurate as $\eta$ increases.

\subsection{Discussion}\label{sec:energy:discussion}

\begin{figure}[t]
\includegraphics{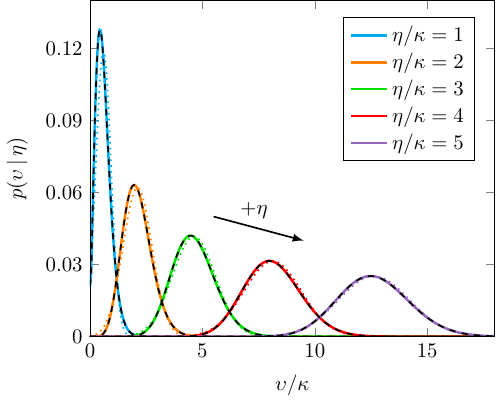}
\caption{\label{fig:energy:probability-fixed-kappa}%
Probability density $p$ of the nondimensional link energy $\upsilon$ given nondimensional force $\eta$ using exact (solid), asymptotic (dashed), and Gaussian (dotted) relations while increasing $\eta$ with fixed nondimensional link stiffness $\kappa=10$.
}
\end{figure}

Similar to the link length in Fig.~\ref{fig:length:average}, the results in Fig.~\ref{fig:energy:average} show that the average link energy is nonzero without an applied force, again due to thermal fluctuations and the tendency toward higher entropy.
Also similar to before, the relative fluctuations in link energy decrease with the applied force $\eta$ as shown in Fig.~\ref{fig:energy:deviation}, and the distribution of link energies narrows and moves toward $\langle\upsilon\rangle\to 0$ as the nondimensional link stiffness $\kappa$ increases as in Fig.~\ref{fig:energy:probability-fixed-eta}.
Dissimilar to the behavior of link lengths, the absolute variance in link energy $\sigma_\upsilon^2$ is sensitive to the applied force, where increasing $\eta$ widens the distribution of link energies as shown in Fig.~\ref{fig:energy:probability-fixed-kappa}.
Further, the link energies do not appear to be nearly normally distributed in most cases, as evidenced by the poor approximation provided by the Gaussian distributions in Figs.~\ref{fig:energy:probability-fixed-eta} and \ref{fig:energy:probability-fixed-kappa}.
Since the link lengths are approximately normally distributed (and are typically one-to-one with the link energy), chain dissociation and similar model criteria are most easily based on the link length for $\kappa\gg 1$.
Nevertheless, the asymptotic relations provided here enable both options.

\section{Error analysis}\label{sec:error}

\begin{figure}[t]
\includegraphics{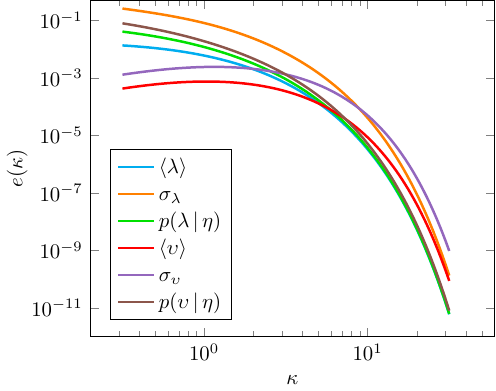}
\caption{\label{fig:error}%
Relative $L_2$ error norm $e(\kappa)$ as a function of the nondimensional link stiffness $\kappa$ for each asymptotic relation compared with the corresponding exact relation.
}
\end{figure}

In order to quantify the accuracy of all the asymptotic relations presented here, the error in each case with respect to the corresponding exact relation is analyzed.
The degree of accuracy is understood both in terms of a decreasing error metric as well as the slope of that error in log-log scale \cite{buche2022freely}.
The relative $L_2$ error norm is applied, so if the approximation error is $\Delta f$ for some quantity $f$, the relative $L_2$ error $e(\kappa)$ is given by
\begin{equation}
e(\kappa) = \sqrt{\frac{\int_0^{\eta_0}\Delta f(\eta)^2\,d\eta}{\int_0^{\eta_0}f(\eta)^2\,d\eta}},
\end{equation}
where $\eta_0 = 10$ is selected.
This value is representative rather than fundamental, defining an interval of forces over which the approximation error is considered.
With link lengths in the range of $\ell_b=0.1$--$1\,\mathrm{nm}$, at room temperature $\eta_0=10$ corresponds to forces in the range of $40$--$400\,\mathrm{pN}$, a range that is relevant to single-molecule force spectroscopy \cite{marko1998dna,radiom2017influence}.
The probability distribution errors are formulated similarly, except both the force and the object of the probability are integrated over,
\begin{equation}
e(\kappa) = \sqrt{\frac{\int_0^{\eta_0}\int_{x^-}^{x^+}\Delta p(x\,|\,\eta)^2\,dx\,d\eta}{\int_0^{\eta_0}\int_{x^-}^{x^+} p(x\,|\,\eta)^2\,dx\,d\eta}},
\end{equation}
where $x^\pm = \langle x\rangle \pm \sigma_x/2$ is selected.
The relative approximation error $e(\kappa)$ is shown in Fig.~\ref{fig:error} as a function of the nondimensional link stiffness $\kappa$ for each of the asymptotic relations derived here: $\langle\lambda\rangle$ in Eq.~\eqref{eq:length:average:asymptotic}, $\sigma_\lambda$ via Eq.~\eqref{eq:length:variance:asymptotic}, $p(\lambda\,|\,\eta)$ in Eq.~\eqref{eq:length:probability:asymptotic}, $\langle\upsilon\rangle$ in Eq.~\eqref{eq:energy:average:asymptotic}, $\sigma_\upsilon$ via Eq.~\eqref{eq:energy:variance:asymptotic}, and finally $p(\upsilon\,|\,\eta)$ via Eqs.~\eqref{eq:energy:probability:exact} and \eqref{eq:length:probability:asymptotic}.
When $\kappa<1$, the error is somewhat considerable in several cases, but this is expected since the derivations assumed $\kappa$ was large.
When $\kappa>1$, the errors begin to decrease, and as $\kappa>10$, the errors all drop dramatically.
Also, since the log-log error slopes do not approach a constant as $\kappa$ increases, where the slope rapidly decreases instead, it is evident that all of these asymptotic relations are accurate to within transcendentally small terms such as $e^{-\kappa/2}$.

Fig.~\ref{fig:error} shows that the asymptotic relations are highly accurate for moderately large stiffness, such as $\kappa\sim 10$, and would become even more accurate as $\kappa$ increases.
Many idealized single-chain models use $\kappa$ values in the hundreds or thousands to match experimental behavior or describe chain dissociation \cite{buche2021chain,mulderrig2023statistical,zhu2025stretching}, so the present approximations should be well within their useful range.
Although the classical freely jointed chain is recovered formally as $\kappa\to\infty$, that inextensible limit is unable to represent the high-force behavior where link stretching dominates, nor can it readily describe link energetics or any dissociation criteria based on link length or energy.
Thus, finite but large $\kappa$ is the practically relevant regime for the present work.

\section{Link dissociation}\label{sec:dissociation}

\begin{figure}[t]
\includegraphics{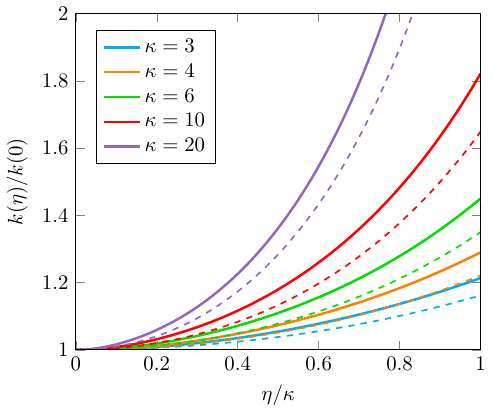}
\caption{\label{fig:dissociation}%
Link dissociation rate $k(\eta)$ as a function of the nondimensional force $\eta$ applied to the chain with (solid) and without (dashed) the effects of link length fluctuations.
}
\end{figure}

Neglecting link length and energy fluctuations can significantly alter predictions of the dissociation for a given link in the chain, which is demonstrated here using a simple model for the dissociation rate: $k(\upsilon)\propto e^{\alpha\upsilon}$, or equivalently, $k(\lambda)\propto e^{\alpha\kappa(\lambda - 1)^2/2}$ \cite{buche2021chain,mulderrig2023statistical}.
This rate is typically evaluated using the average values $\langle\lambda\rangle$ and $\langle\upsilon\rangle$, which with the crudest approximation produces $k(\eta)\propto e^{\alpha\eta^2/2\kappa}$.
Even if the more accurate ensemble averages derived in this work were used instead, such rates would still neglect the contribution of thermal fluctuations.
Instead, the fluctuation-aware dissociation rate of a given link in a chain under force should account for the probability that the link occupies each particular stretch, which is
\begin{equation}\label{eq:dissociation}
k(\eta) = \int_0^\infty k(\lambda)\,p(\lambda\,|\,\eta)\,d\lambda,
\end{equation}
or the equivalent form via $p(\upsilon\,|\,\eta)$.
The preceding approximation $k(\eta) = k[\langle\lambda\rangle]$ corresponds to replacing $p(\lambda\,|\,\eta)$ with the Dirac delta function $\delta[\langle\lambda\rangle - \lambda]$.
This is only valid when $p(\lambda\,|\,\eta)$ is sufficiently narrow relative to the length scale over which $k(\lambda)$ varies.
For the extensible freely jointed chain, this width was shown to be $\sigma_\lambda\sim\kappa^{-1/2}$, and because $k(\lambda)$ depends exponentially on link stretch, the variation of $k(\lambda)$ over this width can remain significant even when $\kappa$ is large.
The link dissociation rate $k(\eta)$ is shown in Fig.~\ref{fig:dissociation} as a function of the nondimensional force $\eta$ applied to the chain, comparing the basic approximation that neglects fluctuations, $k(\eta)=k[\langle\lambda\rangle]$, with the fluctuation-aware relation in Eq.~\eqref{eq:dissociation}, with $\alpha=0.1$.
The basic approximation under-predicts the dissociation rate in all cases, since it does not account for the population of links that fluctuate above the average length, which contributes disproportionately to the rate because $k(\lambda)$ increases exponentially with link energy.

\section{Conclusion}\label{sec:conclusion}

Thermodynamic fluctuations of both link lengths and energies within the extensible freely jointed chain model under an applied force have been analyzed in detail.
New asymptotic relations for the average, variance, and overall probability distribution of both values were derived.
The accuracy of these approximate relations were then verified with respect to the corresponding exact relations, which were also presented analytically.
Specifically, each of these asymptotic relations were shown to be extremely accurate for larger values of the link stiffness.
Since these asymptotic relations are relatively simple to implement, it is recommended that fluctuations in link length or energy should be considered when determining polymer chain dissociation as a function of either value.
Further, since the fluctuations in link length were demonstrated to be approximately normally distributed -- where the corresponding average and variance are easily approximated, and most relevant potential energies governing the links are one-to-one with the link length along the trajectory of separation -- it is recommended that the link length should be selected as the model criterion for predicting chain dissociation.
These results are specific to the isotensional ensemble and likely do not directly apply to the isometric ensemble, where the fixed end-to-end extension introduces correlations between links and makes the resulting statistics more complicated.
Whether the per-link statistics derived here describe link-level fluctuations in a chain under fixed extension remains an open question.
Studying thermodynamic fluctuations of link lengths and energies in the isometric ensemble is therefore an important subject for future work, as is validating the use of the thermodynamic statistics analyzed here for predicting phenomena such as chain dissociation.

\begin{acknowledgments}
Sandia National Laboratories is a multi-mission laboratory managed and operated by National Technology and Engineering Solutions of Sandia, LLC., a wholly owned subsidiary of Honeywell International, Inc., for the U.S. Department of Energy's National Nuclear Security Administration under Contract No. DE-NA0003525. Any subjective views or opinions expressed in the paper do not necessarily represent the views of the U.S. Department of Energy or the U.S. Government. The U.S. Government retains and the publisher, by accepting the article for publication, acknowledges that the U.S. Government retains a nonexclusive, paid-up, irrevocable, world-wide license to publish or reproduce the published form of this manuscript, or allow others to do so, for U.S. Government purposes.
\end{acknowledgments}

\appendix
\section{Exact relations for $\langle\lambda^n\rangle$}\label{sec:appendix}

Using symbolic math tools such as \texttt{Mathematica} \cite{mathematica}, the following exact analytic relations can be obtained.
Significant manipulation of the originally obtained result is required in order to arrive at the forms that follow.
Each of these relations are obtained by evaluating $z_n(\eta)$ in Eq.~\eqref{eq:z:n} analytically for $n=1,\ldots,4$, scaling the result for $z_n(\eta)$ by $z(\eta)$ in Eq.~\eqref{eq:z:exact}, and simplifying the ratio.
For $n=1$, the exact result is
\begin{equation}\label{eq:lambda:1:exact}
\langle\lambda\rangle = \tfrac{\mu_1^+(\kappa,\eta) - \mu_1^-(\kappa,\eta) + \nu_1(\kappa,\eta)}{\mu_0^+(\kappa,\eta) - \mu_0^-(\kappa,\eta)},
\end{equation}
where $\mu_0^\pm(\kappa,\eta)$ is defined as
\begin{equation}\label{eq:mu:pm}
\mu_0^\pm(\kappa,\eta) \equiv e^{\pm\eta} \left(1 \pm \frac{\eta}{\kappa}\right) \left[1 \pm \erf\left(\frac{\eta\pm\kappa}{\sqrt{2\kappa}}\right)\right].
\end{equation}
$\mu_1^\pm(\kappa,\eta)$ is defined as
\begin{equation}
\mu_1^\pm(\kappa,\eta) \equiv e^{\pm\eta} \left[\tfrac{1}{\kappa} + \left(1\pm\tfrac{\eta}{\kappa}\right)^2\right] \left[1 \pm \erf\left(\tfrac{\eta\pm\kappa}{\sqrt{2\kappa}}\right)\right],
\end{equation}
and $\nu_1(\kappa,\eta)$ is defined as
\begin{equation}
\nu_1(\kappa,\eta) \equiv 4\,\tfrac{\eta}{\kappa}\,\tfrac{e^{-\kappa/2}}{\sqrt{2\pi\kappa}}\,e^{-\eta^2/2\kappa}.
\end{equation}
For $n=2$, the exact result is
\begin{equation}\label{eq:lambda:2:exact}
\langle\lambda^2\rangle = \tfrac{\mu_2^+(\kappa,\eta) - \mu_2^-(\kappa,\eta) + \nu_2^+(\kappa,\eta) - \nu_2^-(\kappa,\eta)}{\mu_0^+(\kappa,\eta) - \mu_0^-(\kappa,\eta)},
\end{equation}
where $\mu_2^\pm(\kappa,\eta)$ is defined as
\begin{equation}
\mu_2^\pm(\kappa,\eta) \equiv \left[\tfrac{3}{\kappa} + \left(1\pm\tfrac{\eta}{\kappa}\right)^2\right]\mu_0^\pm(\kappa,\eta),
\end{equation}
and $\nu_2^\pm(\kappa,\eta)$ is defined as
\begin{equation}
\nu_2^\pm(\kappa,\eta) \equiv 2\left[\tfrac{2}{\kappa} + \left(1\pm\tfrac{\eta}{\kappa}\right)^2\right]\,\tfrac{e^{-\kappa/2}}{\sqrt{2\pi\kappa}}\,e^{-\eta^2/2\kappa}.
\end{equation}
For $n=3$, the exact result is
\begin{equation}\label{eq:lambda:3:exact}
\langle\lambda^3\rangle = \tfrac{\mu_3^+(\kappa,\eta) - \mu_3^-(\kappa,\eta) + \nu_3(\kappa,\eta)}{\mu_0^+(\kappa,\eta) - \mu_0^-(\kappa,\eta)},
\end{equation}
where $\mu_3^\pm(\kappa,\eta)$ is defined as
\begin{equation}
\mu_3^\pm(\kappa,\eta) \equiv e^{\pm\eta}\,g_3^\pm(\kappa,\eta) \left[1\pm\erf\left(\tfrac{\eta\pm\kappa}{\sqrt{2\kappa}}\right)\right],
\end{equation}
and $g_3^\pm(\kappa,\eta)$ is defined as
\begin{equation}
g_3^\pm(\kappa,\eta) \equiv \left(1\pm\tfrac{\eta}{\kappa}\right)^4 + \tfrac{6}{\kappa}\left(1\pm\tfrac{\eta}{\kappa}\right)^2 + \tfrac{3}{\kappa^2},
\end{equation}
and $\nu_3(\kappa,\eta)$ is defined as
\begin{equation}
\nu_3(\kappa,\eta) \equiv 4\,\tfrac{\eta}{\kappa^3}\left(\eta^2+5\kappa+3\kappa^2\right)\tfrac{e^{-\kappa/2}}{\sqrt{2\pi\kappa}}\,e^{-\eta^2/(2\kappa)}.
\end{equation}
For $n=4$, the exact result is
\begin{equation}\label{eq:lambda:4:exact}
\langle\lambda^4\rangle = \tfrac{\mu_4^+(\kappa,\eta) - \mu_4^-(\kappa,\eta) + \nu_4(\kappa,\eta)}{\mu_0^+(\kappa,\eta) - \mu_0^-(\kappa,\eta)},
\end{equation}
where $\mu_4^\pm(\kappa,\eta)$ is defined as
\begin{equation}
\mu_4^\pm(\kappa,\eta) \equiv e^{\pm\eta}\,g_4^\pm(\kappa,\eta) \left[1\pm\erf\left(\tfrac{\eta\pm\kappa}{\sqrt{2\kappa}}\right)\right],
\end{equation}
and $g_4^\pm(\kappa,\eta)$ is defined as
\begin{equation}
g_4^\pm(\kappa,\eta) \equiv \left(1\pm\tfrac{\eta}{\kappa}\right)^5 + \tfrac{10}{\kappa}\left(1\pm\tfrac{\eta}{\kappa}\right)^3 + \tfrac{15}{\kappa^2}\left(1\pm\tfrac{\eta}{\kappa}\right),
\end{equation}
and $\nu_4(\kappa,\eta)$ is defined as
\begin{equation}
\nu_4(\kappa,\eta) \equiv \left(\tfrac{16\eta}{\kappa}+\tfrac{72\eta}{\kappa^2}+\tfrac{16\eta^3}{\kappa^3}\right)\tfrac{e^{-\kappa/2}}{\sqrt{2\pi\kappa}}e^{-\eta^2/(2\kappa)}.
\end{equation}

\bibliography{main}

\end{document}